\documentclass[9pt,conference]{IEEEtran}
\IEEEoverridecommandlockouts
\usepackage{cite}
\usepackage{amsmath,amssymb,amsfonts}
\usepackage{algorithmic}
\usepackage{graphicx}
\usepackage{textcomp}
\usepackage{xcolor}
\usepackage{multirow}
\usepackage{makecell}
\usepackage{amsfonts}
\def\BibTeX{{\rm B\kern-.05em{\sc i\kern-.025em b}\kern-.08em
    T\kern-.1667em\lower.7ex\hbox{E}\kern-.125emX}}

\begin{document}

\title{Dynamic Frequency-Adaptive Knowledge Distillation for Speech Enhancement
}

\author{\IEEEauthorblockN{Xihao Yuan}
\IEEEauthorblockA{\textit{Noah's Ark Lab} \\
\textit{Huawei}\\
Shanghai, China \\
yuanxihao@huawei.com}
\\
\IEEEauthorblockN{Lu Zhou}
\IEEEauthorblockA{\textit{Noah's Ark Lab} \\
\textit{Huawei}\\
Shanghai, China \\
zhoulu29@huawei.com}

\and

\IEEEauthorblockN{Siqi Liu}
\IEEEauthorblockA{\textit{Noah's Ark Lab} \\
\textit{Huawei}\\
Shanghai, China \\
liusiqi37@huawei.com}
\\
\IEEEauthorblockN{Jian Li}
\IEEEauthorblockA{\textit{Noah's Ark Lab} \\
\textit{Huawei}\\
Shanghai, China \\
lijian368@huawei.com}

\and

\IEEEauthorblockN{Hanting Chen}
\IEEEauthorblockA{\textit{Noah's Ark Lab} \\
\textit{Huawei}\\
Shanghai, China \\
chenhanting@huawei.com}
\\
\IEEEauthorblockN{Jie Hu}
\IEEEauthorblockA{\textit{Noah's Ark Lab} \\
\textit{Huawei}\\
Shanghai, China \\
hujie23@huawei.com}
}

\maketitle

\begin{abstract}
Deep learning-based speech enhancement (SE) models have recently outperformed traditional techniques, yet their deployment on resource-constrained devices remains challenging due to high computational and memory demands. This paper introduces a novel dynamic frequency-adaptive knowledge distillation (DFKD) approach to effectively compress SE models. Our method dynamically assesses the model's output, distinguishing between high and low-frequency components, and adapts the learning objectives to meet the unique requirements of different frequency bands, capitalizing on the SE task's inherent characteristics. To evaluate the DFKD's efficacy, we conducted experiments on three state-of-the-art models: DCCRN, ConTasNet, and DPTNet. The results demonstrate that our method not only significantly enhances the performance of the compressed model (student model) but also surpasses other logit-based knowledge distillation methods specifically for SE tasks.
\end{abstract}


\section{Introduction}
Deep learning has permeated various aspects of daily life, with significant advancements particularly in the realm of speech enhancement. This field focuses on the suppression of background noise to achieve clearer and more intelligible speech. Recent studies, such as DCCRN\cite{hu2020dccrn}, CRN\cite{tan2018convolutional}, and DEMUCS\cite{defossez2019demucs}, have demonstrated that deep learning algorithms surpass traditional methods in terms of performance. However, the deployment of deep neural networks (DNNs) on devices with limited computational resources poses a significant challenge due to their substantial resource demands. While existing methods attempt to address this issue, they often struggle to strike a balance between maintaining model performance and achieving computational efficiency. Consequently, the development of DNNs that offer effective noise suppression capabilities without incurring excessive computational costs has emerged as a pivotal area of research.
\begin{figure}[t]
  \begin{center}
  \includegraphics[width=\linewidth]{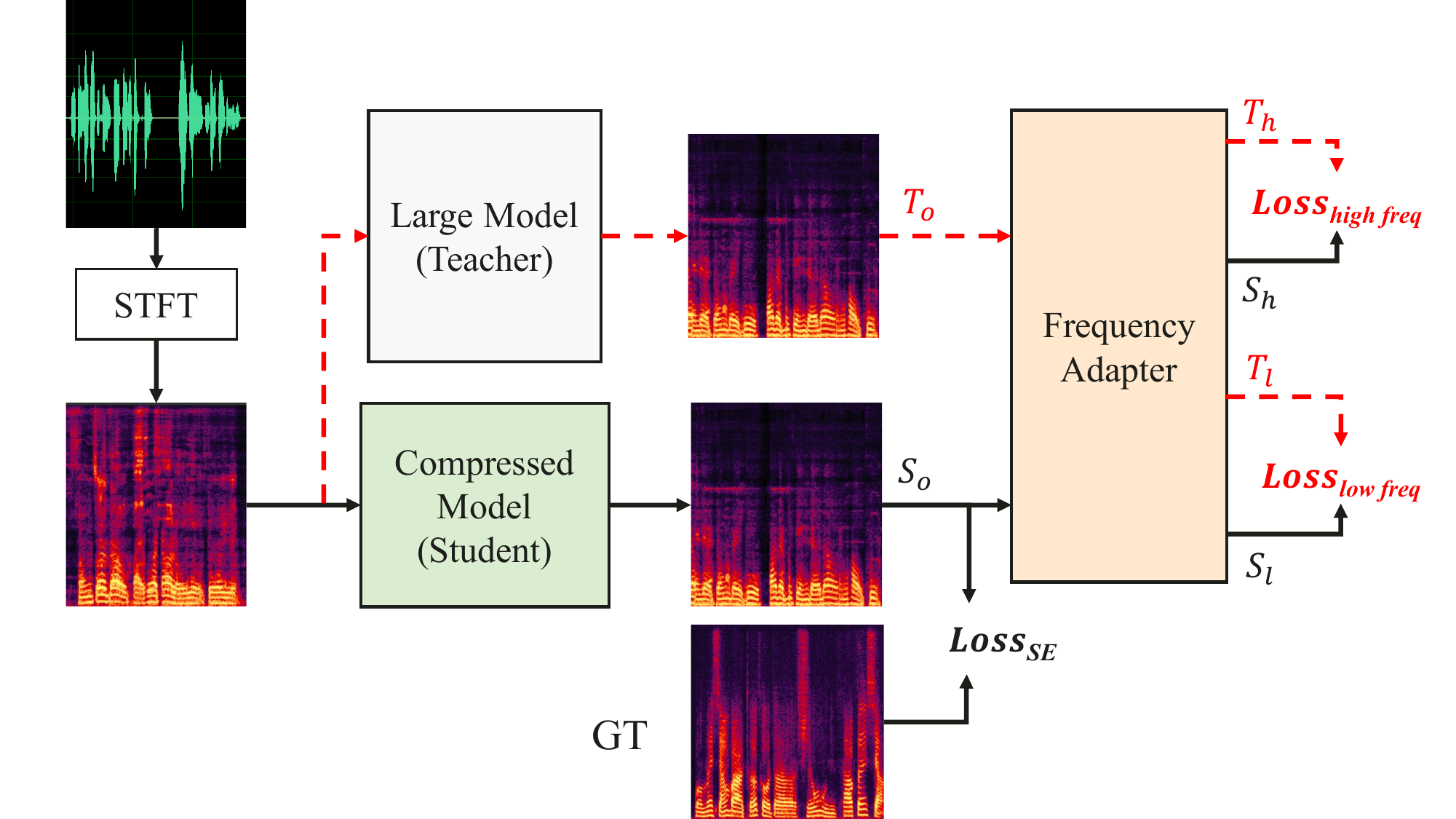}
  \caption{Overview of DFKD. The teacher model's data flow is represented by the red dotted line. The output is separated into high and low-frequency bands using the Frequency Adapter, and different bands are subjected to the specified loss function. $Loss_{SE}$ is a typical loss function in SE tasks, taking on different forms to accommodate the varying architectures of different networks.}
  \label{fig:DFKD_overview}
  \end{center}
\end{figure}

The primary strategies for model compression include pruning, tensor decomposition, quantization, and knowledge distillation (KD). Pruning, which capitalizes on the inherent redundancy in model parameters, targets the elimination of less critical parameters to streamline the model\cite{liu2017learning}\cite{han2015learning}\cite{dong2017learning}\cite{xiao2019autoprune}. Quantization, conversely, focuses on reducing the bit depth of parameters, typically replacing 16 or 32-bit values with lower-bit alternatives\cite{banner2018scalable}\cite{chmiel2020neural}\cite{courbariaux2014training}. Tensor decomposition, on the other hand, employs matrix or convolutional factorization techniques to pinpoint and discard non-essential parameters, further optimizing model size and complexity\cite{de2000multilinear}. Unlike the previous three model compression methods, which often necessitate specialized hardware, KD offers a more flexible approach. It involves compressing a large, complex teacher model into a smaller, more efficient one, with minimal constraints on the design of the compressed model. This paper primarily focuses on the application of KD in the domain of speech enhancement.

Initially introduced by Hinton in 2015\cite{hinton2015distilling}, KD is a process that facilitates the transfer of knowledge from a larger model to a smaller one. By doing so, it enables the student model to emulate the performance characteristics of the teacher model, thereby enhancing its capabilities on specific tasks. KD can be broadly classified into three categories: logits-based, feature-based, and relation-based techniques. Logits-based KD involves calculating the KD loss by comparing the outputs of the teacher and student models. This approach has been extensively researched and has found practical applications\cite{hinton2015distilling}\cite{kim2017transferring}\cite{ding2019adaptive}. Feature-based KD, such as FT\cite{kim2018paraphrasing} and DFA\cite{guan2020differentiable}, extend the traditional KD approach by considering not only the outputs but also the intermediate features of the teacher and student models. When the student model differs structurally from the teacher, an auxiliary structure may be introduced to align the feature dimensions, enabling the computation of a tailored loss\cite{heo2019knowledge}\cite{zagoruyko2016paying}\cite{heo2019comprehensive}.

In contrast to these, relation-based KD delves into the relationships between feature maps across different layers. Techniques such as FSP\cite{yim2017gift}, CCKD\cite{peng2019correlation}, and LP\cite{chen2020learning} aim to efficiently construct a meaningful relationship matrix that captures the inter-layer dependencies between the teacher and student models.

While KD has made significant strides in computer vision (CV) and natural language processing (NLP), its application in SE remains largely uncharted.  Existing KD approaches \cite{hao2020sub}\cite{wang2022text}\cite{thakker2022fast}\cite{chen2021cross}\cite{wan2023abc} for SE often overlook the distinct noise characteristics of high and low frequency bands by directly computing the loss and distilling the network's output. Other frequency-differentiating approaches, like Suband-KD\cite{hao2020sub}, rely on empirically determined fixed crossover frequencies. This practice fails to account for the diverse and intricate frequency distributions inherent in various acoustic environments, particularly in the nuanced treatment of high and low frequency components.

In this paper, we introduce a dynamic frequency-adaptive knowledge distillation method designed to bridge this gap. Our technique discerns between high and low-frequency bands within speech signals across diverse conditions, such as varying noise levels and distinctions between male and female voices. By tailoring the loss functions to the characteristics of these frequency bands, we guide the student model to achieve a more nuanced balance in noise suppression and speech enhancement. This adaptive approach to KD in SE represents a significant advancement over existing methods, offering improved performance in the compression and enhancement of speech signals.

The rest of this paper is organized as follows. In Section 2, we introduce the intricacies of our proposed method. Section 3 elaborates on the experimental details including dataset, implementation and the analysis of the results. Concluding remarks and implications for future research are discussed in Section 5.

\section{Methodology}

\subsection{System Overview}

DFKD operates on frequency-domain inputs, as illustrated in Figure 1. We initiate the process by transforming the original time-domain audio signals into the frequency domain using a Short Time Fourier Transform (STFT). The resultant frequency-domain signals are then fed into both the teacher and student models, yielding the teacher output $T_o$ and the student output $S_o$, respectively. 

Furthermore, our Frequency Adapter module performs real-time analysis of the frequency-domain signals from $T_o$ and $S_o$, identifying the crossover points that delineate high and low frequencies. This module effectively bifurcates the signals into high-frequency components $T_h$ and $S_h$, and low-frequency components $T_l$ and $S_l$.

In the training phase, we employ three distinct loss functions: $L_{SE}$, which calculates loss between $S_o$ and the ground truth (GT). Following the original paper, it takes different forms for different networks, such as scale-invariant source-to-noise ratio (SI-SNR) loss for DCCRN; $L_{high freq}$, which quantifies the discrepancy between the student and teacher models' high-frequency outputs; and $L_{low freq}$ , which assesses the difference in low-frequency outputs. Throughout the DFKD training, the teacher model's parameters are kept constant, while the student model's parameters are iteratively updated based on these three loss functions.

\subsection{Frequency Adapter}

\begin{figure}[t]
  \begin{center}
  \includegraphics[width=\linewidth]{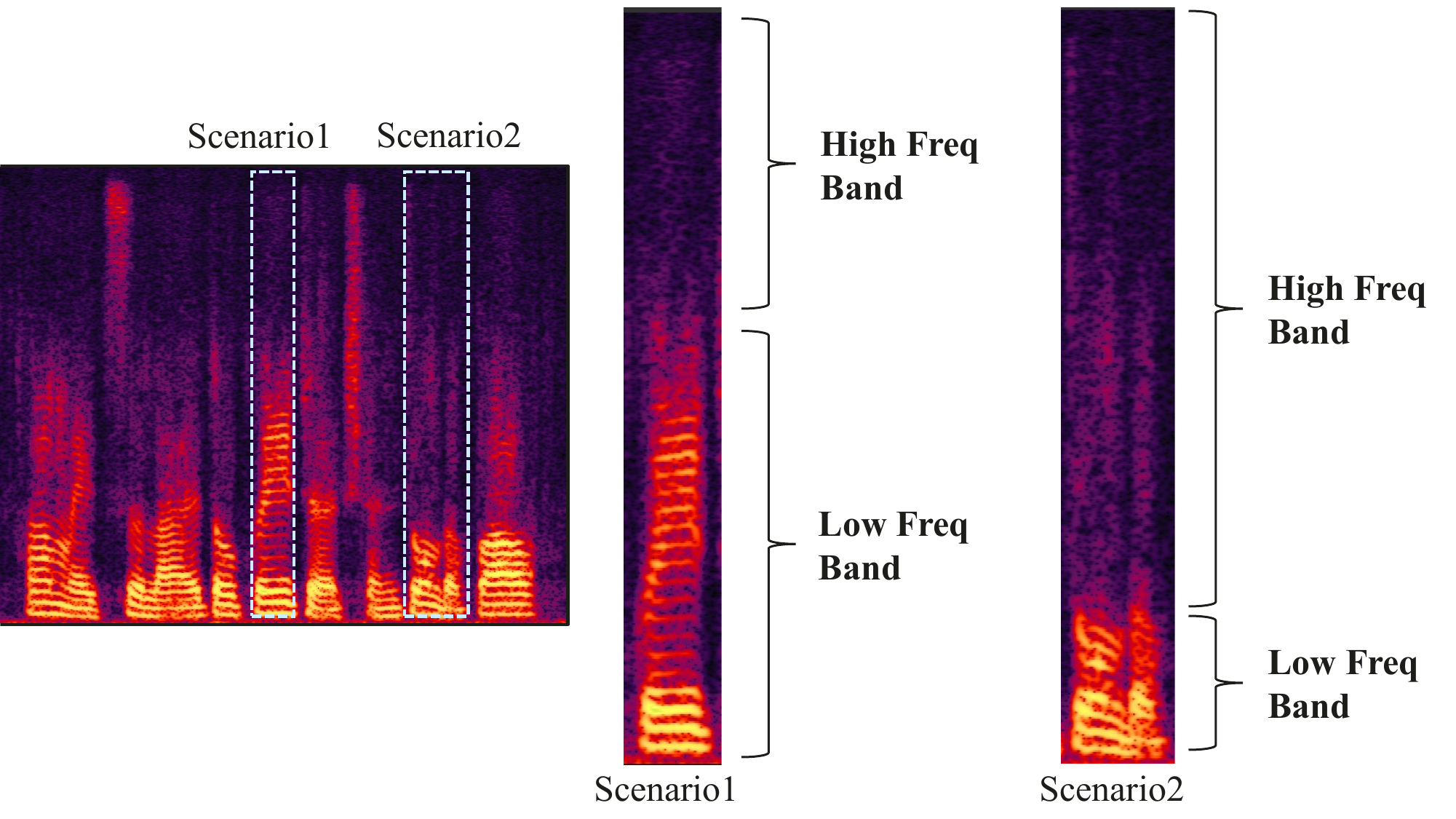}
  \caption{Different Scenarios for Frequency Adapter. Frequency Adapter senses the characteristics of a scene and adaptively divides the frequency bands, as the distribution of high and low frequencies of human voice changes over time within one speech clip.}
  \label{fig:Frequency_Adapter}
  \end{center}
\end{figure}

As illustrated in Figure 2, the frequency spectra in real-world scenarios are inherently complex and the distribution of speech and noise can fluctuate significantly within short time segments. Consequently, KD methods that ignore the variable characteristics of high and low frequency components or rely on fixed crossover points can be detrimental to a model's adaptability. These methods often fail to account for the unique noise characteristics present in different frequency bands, which is a critical factor for effective speech enhancement.

The Frequency Adapter module plays a pivotal role in DFKD framework by identifying the optimal crossover points between high and low frequency bands in real-time. It dynamically segments the network's output signal into two distinct frequency bands and tailors the optimization of these bands using frequency-specific loss functions. This targeted approach refines the student model's performance, ensuring that both speech retention and noise suppression capabilities are optimized.

In more detail, let's consider the output of the teacher model, denoted as $T_o$. For each sequence segment ranging from 0 to 256 (assuming that we obtain 257 frequency bins after STFT), the module calculates the maximum value $F_o$ from $T_o$. Here, $f_i$ represents the maximum value observed from $t_0$ to $t_i$ within the sequence. This process allows for the adaptive determination of frequency band boundaries that are most relevant for the given audio signal, facilitating a more nuanced and effective distillation process.

\begin{equation}
    T_o = (t_0, t_1, t_2, ..., t_{256})
\end{equation}

\begin{equation}
\begin{aligned}
    F_o = (f_0, f_1, f_2, ..., f_{256}), \text{where}\  f_i = \text{max}(t_0, t_1, t_2, ..., t_i)
\end{aligned}
\end{equation}

Following the acquisition of the sequence $F_o$ , we proceed to compute its first-order derivative at each frequency point. To mitigate the risk of division by zero errors during computation, an infinitesimal value $\epsilon$ is incorporated. This precaution ensures numerical stability in our calculations. Subsequently, we identify the position $m$ at which the first-order derivative attains its maximum value. This position $m$ serves as the demarcation point, enabling us to segment the outputs of both the teacher model and the student model into high-frequency and low-frequency components.

\begin{equation}
\begin{aligned}
    \nabla F_o = (\nabla f_0, \nabla f_1, \nabla f_2, ..., \nabla f_{255}), \\ \text{where}\ \nabla f_i = \frac{f_{i+1} - f_i}{f_i + \epsilon}\ (i\in[0, 255])
\end{aligned}
\end{equation}

\begin{equation}
    m = \text{argmax}(\nabla F_o)
\end{equation}

\begin{equation}
\begin{aligned}
    &S_h = (s_0, s_1, s_2, ..., s_m), 
    &S_l = (s_m, s_{m+1}, ..., s_{256})
\end{aligned}
\end{equation}

\begin{equation}
\begin{aligned}
    &T_h = (t_0, t_1, t_2, ..., t_m), 
    &T_l = (t_m, t_{m+1}, ..., t_{256})
\end{aligned}
\end{equation}

The high-frequency ($T _h$) and low-frequency ($T_l$) components of the teacher's output, along with the corresponding high-frequency ($S_h$) and low-frequency ($S_l$) components of the student's output, are utilized to apply distinct loss functions tailored to each frequency band. This strategic approach is designed to enhance the student's full-band noise reduction capabilities. By leveraging frequency-specific loss functions, we aim to refine the student's performance in noise suppression and speech quality enhancement across the entire spectrum of frequencies.

\subsection{KD Loss}

In the field of speech enhancement, achieving a delicate balance between noise suppression and speech retention is paramount. While aggressive noise suppression may inadvertently diminish the clarity of speech by reducing vocal timbre, an excessive focus on speech retention can result in inadequate noise reduction, thereby degrading the listening experience. The ideal model for SE must effectively suppress noise without compromising the quality and integrity of the speech signal, particularly within the typical human speech frequency range of 1-4 kHz. Below 4 kHz, the spectrum is a complex blend of speech and noise, where speech is more dominant. Above 4 kHz, noise tends to be more prevalent. However, the distinction between high and low frequencies is not strictly bounded by the 4 kHz marker, as it varies with acoustic context, including differences in male and female voice frequency ranges. Therefore, an SE model must be adaptable to these variations to optimize performance.

Leveraging the understanding that different frequency bands exhibit distinct distributions of noise and speech, we can tailor the guidance provided by loss functions to each band. This approach allows for a more nuanced optimization of the model, where the high-frequency and low-frequency components are treated differently to better preserve speech quality and effectively suppress noise.

In scenarios where speech and noise are mixed in the low-frequency range, our objective is to develop a model that effectively separates speech while maximizing noise suppression. Conventional approaches often aim to simultaneously regulate both amplitude and phase. However, this coupled approach to amplitude and phase optimization can lead to suboptimal outcomes for both speech and noise components.

Given the phase invariance characteristic of speech signals and the arbitrary increase in amplitude, our method modifies this approach. We relax the amplitude constraints and introduce an enhanced cosine loss for the low-frequency band. This adaptation encourages the model to prioritize phase constraints, thereby improving its ability to distinguish between speech and noise.

The original cosine loss, which has a value range of $[-1, 1]$, can cause issues with gradient anomalies and reversed update directions during the backpropagation phase of training. To address this, we adjust the cosine loss by subtracting 1, shifting its range to $[-2, 0]$. This modification ensures a consistent update direction for the network, facilitating more effective training and optimization.

\begin{equation}
\begin{aligned}
    L_{\text{cosine}}(T_o, S_o) = cos(T_o, S_o) - 1
\end{aligned}
\end{equation}

\begin{equation}
\begin{aligned}
    L_{\text{low freq}} = L_{\text{cosine}}(T_l, S_l)
\end{aligned}
\end{equation}

The high-frequency segment predominantly consists of noise, thus the primary objective in this frequency range is noise suppression, with a particular focus on amplitude constraints. However, in certain scenarios, this frequency band may also contain subtle vocal timbres essential for speech quality. Consequently, phase constraints are necessary to preserve these speech characteristics. To address this dual requirement, we employ a combined loss function that incorporates both $L_{\text{l2}}$ for amplitude constraints and $L_{\text{cosine}}$ for phase constraints. Additionally, we introduce a hyperparameter $\beta$ during training, which serves as a control mechanism. This parameter allows us to adjust the model's emphasis between noise suppression and speech retention, ensuring a balanced approach tailored to the specific requirements of the scenario.

\begin{equation}
\begin{aligned}
    L_{\text{high freq}} = & \beta * L_{\text{cosine}}(T_h, S_h) + (1 - \beta) * L_{\text{l2}}(T_h, S_h)
\end{aligned}
\end{equation}

\begin{equation}
\begin{aligned}
    L_{\text{kd}} = L_{\text{low freq}} + L_{\text{high freq}}
\end{aligned}
\end{equation}

\begin{equation}
\begin{aligned}
    L_{\text{total}} = \alpha * L_{\text{kd}} + (1 - \alpha) * L_{\text{SE}}
\end{aligned}
\end{equation}

Finally, we adhere to the conventional logits-based KD approach to achieve a balance between the knowledge distillation loss ($L_{\text{kd}}$) and $L_{SE}$. This balance is modulated by the hyperparameter $\alpha$. It is essential to re-emphasize that the formulation of $L_{SE}$ is meticulously aligned with the original papers for the respective models.

\section{Experiments}

\subsection{Datasets}

The experimental models in our study were trained using a dataset synthesized from the DNS2020 challenge\cite{reddy2020interspeech}. This comprehensive dataset comprises 500 hours of pristine speech from 2150 unique speakers, along with 65000 noise clips spanning 150 distinct audio classes sourced from Audioset, Freesound, and YouTube videos. To prepare the dataset for training, each audio clip was segmented into 6-second intervals, and the entire training set was uniformly resampled at a rate of 16 kHz to ensure consistency across all data points. The signal-to-noise ratio (SNR) levels were randomly sampled between 0 and 20 dB according to a uniform distribution.

To substantiate the efficacy of our algorithm, we conducted a comprehensive evaluation across various test sets. This included the dev-test subset from the DNS2020 challenge and the combined VoiceBank+DEMAND test set\cite{valentini2016investigating}. Consistency in sampling frequency between the training and testing phases was ensured by resampling all test clips to 16 kHz, aligning with the training set's specifications.

\subsection{Implementation Details}

To evaluate the generalizability of DFKD, we performed experiments on diverse models including DCCRN-CL\cite{hu2020dccrn}, ConvTasNet\cite{luo2019conv}, and DPTNet\cite{chen2020dual}, each with unique architectural designs. For instance, ConvTasNet primarily operates in the time domain, whereas DCCRN-CL is adept at handling both time and frequency domains.

Our experiments compared DFKD's performance against traditional logits-based knowledge distillation methods: L1, L2, KL, ABC-KD, and Suband-KD. The L1, L2, and KL methods follow the conventional knowledge distillation framework by Hinton et al.\cite{hinton2015distilling}, substituting the knowledge distillation loss $L_{kd}$ with $L_{l1}$, $L_{l2}$, and $L_{kl}$, respectively. ABC-KD and Suband-KD are recognized for their efficacy in speech enhancement tasks. Notably, to maintain fair experimental conditions, we used the complete set of labeled training data for ABC-KD, diverging from its original 50\% data usage.

As detailed in Table 1, we further created two student model configurations by reducing the channels of the teacher model, achieving 45\% (small) and 75\% (tiny) reductions in Floating Point Operations (FLOPs) relative to the teacher model. To ensure the accuracy of our experiments, we retrained the teacher models for each configuration following the original papers' guidelines meticulously. The student models were then trained using the same hyperparameters as their corresponding teacher models, establishing a baseline for comparison.

\begin{table}[!tbp]
\caption{FLOPs and Parameters of Models in our experiments.}
\begin{center}
\resizebox{7.5cm}{!}{
\centering

\begin{tabular}{ccc}
\hline
\textbf{Model} & \textbf{FLOPs(G)} & \textbf{Params(M)} \\

\hline

\makecell{DCCRN-CL(teacher)} & 13.47 & 14.66 \\
\makecell{DCCRN-CL-small} & \makecell{7.45 (\textit{-45\%})} &  \makecell{8.77 (\textit{-40\%})} \\
\makecell{DCCRN-CL-tiny} & \makecell{3.33 (\textit{-75\%})} & \makecell{4.26 (\textit{-71\%})} \\

\hline

\makecell{ConvTasNet(teacher)} & 4.91 & 19.67 \\
\makecell{ConvTasNet-small} & \makecell{2.52 (\textit{-49\%})} & \makecell{10.11 (\textit{-49\%})} \\
\makecell{ConvTasNet-tiny} & \makecell{1.32 (\textit{-73\%})} & \makecell{5.33 (\textit{-73\%})} \\

\hline

\makecell{DPTNet(teacher)} & 11.73 & 11.13 \\
\makecell{DPTNet-small} & \makecell{6.63 (\textit{-44\%})} & \makecell{6.27 (\textit{-44\%})} \\
\makecell{DPTNet-tiny} & \makecell{2.97 (\textit{-75\%})} & \makecell{2.8 (\textit{-75\%})} \\

\hline

\end{tabular}
}
\end{center}
\end{table}

\subsection{Experimental Results and Discussion}

Our experimental results, as summarized in Table 2, highlight the superior performance of our method across various model configurations. Notably, DFKD outperforms traditional distillation techniques and even the teacher model under the DCCRN-CL-small and DCCRN-CL-Tiny setups. The teacher model, with 1.8X and 4X the FLOPs of the small and tiny models respectively, was surpassed by the small model's PESQ score by 0.04 points, while the tiny model matched the teacher's performance. In contrast, other distillation methods, such as Suband-KD, showed a significant decrease compared to their baselines, further emphasizing DFKD's effectiveness.

In the case of the DPTNet model, DFKD again led to the most significant improvements, with the small model's PESQ score increasing by 0.026 and the tiny model by 0.112 over their baselines. Suband-KD also showed improvements, particularly for the tiny model, with a 0.079 point increase over the baseline.

For the ConvTasNet model, which operates in the time domain without LSTM units and is thus incompatible with ABC-KD and Suband-KD, DFKD still achieved optimal results. The ConvTasNet-small model's PESQ score improved by 0.081 points, outperforming the teacher model with 1.9X more parameters. These findings underscore DFKD's ability to enhance model performance regardless of the model's architecture or domain of operation.

The experimental outcomes underscore the superiority of our method over other distillation techniques across all evaluated models. The performance variability observed in other distillation methods across different domains or models highlights the robustness and reliability of DFKD. These results substantiate the effectiveness and generalizability of our approach, indicating its potential to consistently enhance model performance irrespective of the underlying model architecture or domain-specific challenges.

In addition to our primary experiments, we conducted extensive cross-dataset testing utilizing the VoiceBank+DEMAND dataset with models initially trained on the DNS2020. Our method not only maintained its state-of-the-art (SOTA) performance but also demonstrated the potential for the student models to surpass their teacher counterparts in specific configurations. Notably, the student models derived from DCCRN-CL-small/tiny, DPTNet-small/tiny, and ConvTasNet-small exhibited superior performance when distilled using DFKD, underscoring the robustness and generalizability of our approach across different datasets.

\begin{table}[!tbp]
\begin{center}

\caption{Comparison of PESQ between different KD methods.}
\resizebox{8.7cm}{!}{
\centering
\begin{tabular}{cccccc}
\hline
\multirow{3}*{\textbf{Model}} & \multirow{3}*{\textbf{Methods}} & \multicolumn{4}{c}{\textbf{Test Dataset}} \\
\multicolumn{2}{c}{} & \multicolumn{2}{c}{\textbf{DNS2020-test}} & \multicolumn{2}{c}{\textbf{VoiceBank+DEMAND}} \\
\multicolumn{2}{c}{} & \makecell{\textbf{PESQ}} & \makecell{\textbf{STOI(\%)}} & \makecell{\textbf{PESQ}} & \makecell{\textbf{STOI(\%)}} \\

\hline

\makecell{DCCRN-CL\\(teacher)} & Scratch & 3.22 & 94.89 & 3.389 & 91.79 \\
\hline
    
\multirow{8}*{\makecell{DCCRN-CL-small}} & Scratch & 3.206 & 94.94 & 3.387 & 91.88 \\
\multicolumn{1}{c}{} & L1 & \textcolor{blue}{3.215} & \textcolor{blue}{94.97} & 3.38 & 91.69 \\
\multicolumn{1}{c}{} & L2 & 3.207 & 94 & 3.389 & 91.92 \\
\multicolumn{1}{c}{} & KL & 3.2 & 94.85 & 3.404 & 92.07 \\
\multicolumn{1}{c}{} & ABC-KD & 3.214 & 94.95 & 3.382 & 91.87 \\
\multicolumn{1}{c}{} & Suband-KD & 3.18 & 94.69 & \textcolor{blue}{3.417} & \textcolor{red}{92.53} \\
\multicolumn{1}{c}{} & DFKD(ours) & \textcolor{red}{3.262} & \textcolor{red}{95.13} & \textcolor{red}{3.431} & \textcolor{blue}{92.2}\\
\hline

\multirow{8}*{\makecell{DCCRN-CL-tiny}} & Scratch & 3.171 & 94.62 & 3.385 & 91.74 \\
\multicolumn{1}{c}{} & L1 & 3.179 & 94.73 & 3.387 & 91.78 \\
\multicolumn{1}{c}{} & L2 & 3.18 & 94.7 & \textcolor{blue}{3.39} & 91.91 \\
\multicolumn{1}{c}{} & KL & \textcolor{blue}{3.19} & \textcolor{blue}{94.81} & \textcolor{blue}{3.39} & \textcolor{red}{92.25} \\
\multicolumn{1}{c}{} & ABC-KD & 3.175 & 94.4 & 3.357 & 91.67 \\
\multicolumn{1}{c}{} & Suband-KD & 3.142 & 94.58 & 3.374 & 92.14 \\
\multicolumn{1}{c}{} & DFKD(ours) & \textcolor{red}{3.224} & \textcolor{red}{94.9} & \textcolor{red}{3.417} & \textcolor{blue}{92.16} \\

\hline

\makecell{DPTNet\\(teacher)} & Scratch & 3.294 & 96.14 & 3.309 & 93.13 \\
\hline

\multirow{8}*{\makecell{DPTNet-small}} & Scratch & 3.255 & \textcolor{red}{93.7} & 3.306 & \textcolor{blue}{89.31} \\
\multicolumn{1}{c}{} & L1 & 3.274 & 92.04 & 3.312 & 88\\
\multicolumn{1}{c}{} & L2 & \textcolor{blue}{3.278} & 90.52 & 3.304 & 86.47 \\
\multicolumn{1}{c}{} & KL & 3.275 & 90 & \textcolor{blue}{3.315} & \textcolor{red}{90.06} \\
\multicolumn{1}{c}{} & ABC-KD & 2.878 & 91.59 & 3.092 & 85.32 \\
\multicolumn{1}{c}{} & Suband-KD & 2.965 & 88.63 & 3.181 & 78.96 \\
\multicolumn{1}{c}{} & DFKD(ours) & \textcolor{red}{3.281} & \textcolor{blue}{92.05} & \textcolor{red}{3.338} & 86.12 \\
\hline

\multirow{8}*{\makecell{DPTNet-tiny}} & Scratch & 3.105 & 95.02 & 3.241 & \textcolor{red}{92.7} \\
\multicolumn{1}{c}{} & L1 & 3.163 & 95.28 & 3.261 & 92.63 \\
\multicolumn{1}{c}{} & L2 & 3.171 & 95.28 & 3.262 & 92.5 \\
\multicolumn{1}{c}{} & KL & 3.158 & 95.27 & 3.254 & \textcolor{blue}{92.69} \\
\multicolumn{1}{c}{} & ABC-KD & 3.013 & 94.57 & 3.205 & 92.3 \\
\multicolumn{1}{c}{} & Suband-KD & \textcolor{blue}{3.184} & \textcolor{blue}{95.3} & \textcolor{blue}{3.298} & 92.65 \\
\multicolumn{1}{c}{} & DFKD(ours) & \textcolor{red}{3.217} & \textcolor{red}{95.32} & \textcolor{red}{3.31} & \textcolor{red}{92.7}\\

\hline

\makecell{ConvTasNet\\(teacher)} & Scratch & 3.364 & 96.35 & 3.425 & 93.25 \\
\hline

\multirow{8}*{\makecell{ConvTasNet-small}} & Scratch & \textcolor{blue}{3.314} & 96.21 & \textcolor{blue}{3.42} & 93.33 \\
\multicolumn{1}{c}{} & L1 & 3.145 & \textcolor{blue}{96.29} & 3.396 & \textcolor{blue}{93.37} \\
\multicolumn{1}{c}{} & L2 & 3.254 & 96.2 & 3.364 & 93.36 \\
\multicolumn{1}{c}{} & KL & 3.105 & 95.2 & 3.236 & 92.79 \\
\multicolumn{1}{c}{} & ABC-KD & - & - & - & -\\
\multicolumn{1}{c}{} & Suband-KD & - & - & - & -\\
\multicolumn{1}{c}{} & DFKD(ours) & \textcolor{red}{3.345} & \textcolor{red}{96.42} & \textcolor{red}{3.463} & \textcolor{red}{93.51} \\
\hline

\multirow{8}*{\makecell{ConvTasNet-tiny}} & Scratch & \textcolor{blue}{3.256} & \textcolor{blue}{95.83} & 3.364 & 92.86\\
\multicolumn{1}{c}{} & L1 & 3.251 & 95.78 & \textcolor{blue}{3.368} & \textcolor{blue}{93.01} \\
\multicolumn{1}{c}{} & L2 & 3.251 & 95.78 & 3.349 & 92.8 \\
\multicolumn{1}{c}{} & KL & 3.148 & 94.9 & 3.273 & 92.61 \\
\multicolumn{1}{c}{} & ABC-KD & - & - & - & - \\
\multicolumn{1}{c}{} & Suband-KD & - & - & - & -\\
\multicolumn{1}{c}{} & DFKD(ours) & \textcolor{red}{3.291} & \textcolor{red}{96.03} & \textcolor{red}{3.394} & \textcolor{red}{93.22} \\

\hline

\end{tabular}
}
\end{center}
\end{table}

\section{Conclusions}

This study introduces an innovative dynamic frequency-adaptive knowledge distillation method tailored for SE tasks. Our method dynamically segments frequency bands in real-time according to scene-specific characteristics and selectively applies tailored loss functions to these bands, representing a novel contribution to knowledge distillation. It has demonstrated superior efficacy, surpassing traditional logits-based techniques and, in some cases, outperforming teacher models with greater computational demands. Extensive experiments across various datasets and models substantiate the method's wide generalizability, highlighting its capability to improve SE in multiple acoustic contexts and network architectures.

\newpage

\bibliographystyle{IEEEtran}
\bibliography{mybib}

\end{document}